\documentclass[preprint,12pt,3p]{elsarticle}

\usepackage{amssymb}
\usepackage[latin1]{inputenc}
\usepackage{amsmath}
\usepackage{graphicx}
\usepackage{placeins}
\usepackage{caption}
\usepackage{subcaption}
\usepackage{pdfpages}
\usepackage{simplewick}

\journal{Nuclear Physics B}

\begin{document}

\begin{frontmatter}

\title{Generalized Additivity in Unitary Conformal Field Theories}

\author[label1,label2]{Gideon Vos\corref{cor1}\fnref{label3}}
\address[label1]{Van Swinderen Institute for Particle Physics and Gravity, University of Groningen,\\Nijenborgh 4, 9747 AG Groningen, The Netherlands.\\This article is registered under preprint number: /hep-th/1411.7941}


\ead{g.vos@rug.nl}



\begin{abstract}
It was demonstrated in \cite{FKPSD},\cite{FKcoef} that $d=4$ unitary CFT's satisfy a special property: if a scalar operator with conformal dimension $\Delta$ exists in the operator spectrum, then the conformal bootstrap demands that large spin primary operators have to exist in the operator spectrum of the CFT with a conformal twist close to $2\Delta+2N$ for any integer $N$. In this paper the conformal bootstrap methods in \cite{KZ} that were used to find the anomalous dimension of the $N=0$ operators have been generalized to recursively find the anomalous dimension of all large spin operators of this class. In AdS these operators can be interpreted as the excited states of the product states of objects that were found in other works.
\end{abstract}


\end{frontmatter}


\section{Introduction}
Over the last 40 years there has been a large interest in conformal field theories (CFTs'). This class of theories is both interesting from a mathematical point of view as well as from a physical point of view. Mathematically the Coleman-Mandula theorem ensures that the conformal group is the largest space-time symmetry that a non-trivial non-supersymmetric theory can posses, while the symmetry itself completely fixes the two- and three-point functions up to normalization. Physically they describe a variety of models, such as statistical systems near critical points and theories at the fixed points of renormalization group-flow. 

For the purpose of this paper the main point of interest as to why these theories are worth studying is their application to AdS/CFT, i.e. a duality between type IIb string theory in the bulk of AdS and $\mathcal{N}=4$ SYM on the boundary \cite{Maldacena}.


Since conformal transformations can be applied to rotate any set of four points onto a plane it has always been natural for CFT's (that live in a space that contains at least two spatial dimension) to map four points onto an Euclidean plane to create an essentially Euclidian theory. A while ago though it was demonstrated in \cite{KZ}\cite{FKPSD} that by considering CFT's on a Minkowski background an interesting property reveals itself. By assuming only that a CFT satisfies unitarity (ie. All OPE coefficients are positive definite and the conformal dimension of local operators satisfy a unitarity bound.) and a crossing symmetry, they have demonstrated that any CFT contains a sector of operators with large spin that behaves as a generalized free theory (GFT). Furthermore they have explicitly calculated the anomalous dimension associated with these large spin operators that contribute to leading order.

In this paper their analysis has been expanded to the subleading terms in the conformal partial wave expansion of the s-channel four-point function. The result is that as was predicted in \cite{KZ} \cite{FKPSD} that not only do operators have to exist with a conformal twist that lies in a region around $2\Delta$ but there actually has to exist a tower of limit points in twist space at $2\Delta+2N$ for any integer $N$. This supports the conclusion that every CFT has a region which behaves as a GFT. From this the conclusion can be drawn that applying the AdS/CFT dictionary to any CFT results in a theory which has a high spin region that can be interpreted as a Fock space, which implies that that region can be interpreted as a gravitational particle theory in the weak coupling limit.

Beyond that the main focus of this paper is the anomalous dimensions associated with these higher order conformal blocks. These have been calculated and have been shown to just as for the $N=0$ case to be inversely proportional to spin. The interpretation that the large spin limit in AdS behaves as a free field theory stays intact. Besides that it is shown that there is a good correspondence between the derived expression for the anomalous dimension and the dual shock wave calculation in AdS done in \cite{penedones1},\cite{penedones2}.

This paper has been divided in the following way. In section 2 a short review will be given on the conformal bootstrap. It exists mainly to establish the notation that will be used throughout this paper. In section 3 a summary will be given of the methods and results obtained in \cite{KZ}. Section 4 will contain an argument to the existence of operators with conformal dimension close to $2\Delta+2N$ and contains the deriviation of a recursion relation for the anomalous dimensions of these double trace operators. Section 5 will contain some arguments from AdS that demonstrates that these anomalous dimensions reproduce the right leading order behaviour if the conformal dimension $\Delta$ of the operators that appear in the four-point function is large.  

\section{The operator product expansion}
The operator product expansion is the most important conformal field theory technique applied in this paper, as all results obtained are derived from this principle. The point of this section will be to establish the notation that will be used throughout this paper (for a short review see section 3 of \cite{RattazziPaper}). The operators product expansion is the statement that any bilocal product of operators can be decomposed into a sum over all local operators in the spectrum, i.e.
\begin{equation}
O_1(x)O_2(0)=\sum_{O^{(k)}} \frac{c_{12k}}{|x|^{\Delta_1+\Delta_2-\Delta_k}}\frac{x^{\mu_1}...x^{\mu_s}}{|x|^s}O^{(k)}_{\mu_1...\mu_s}(x)
\label{OPE}
\end{equation}
Note that it is implied that this decomposition only makes sense when both the left and right hand side are located within an $n$-point function. Typically the two-point functions are diagonalized with respect to the conformal primaries and the non-zero coefficients are set to 1. In this case consistency with the three-point function demands the coefficient that appears in the OPE is the same as the one that appears in the three-point functions.

Since conformal invariance fixes the two and three-point functions and since we can apply the OPE to reduce any $n$-point function to a $(n-1)$-point function we know everything there is to know about a CFT if we know all of the operators in the spectrum and all of their respective OPE coefficients.

\subsection{Conformal partial wave decomposition}
Unlike the two- and three-point function the four-point function is not fixed by symmetry. The reason for this is that for four points in spacetime we can construct two conformal invariants
\begin{equation}
\begin{array}{ll}
u=\frac{(x_1-x_2)^2(x_3-x_4)^2}{(x_1-x_3)^2(x_2-x_4)^2}\,,&v=\frac{(x_1-x_4)^2(x_2-x_3)^2}{(x_1-x_3)^2(x_2-x_4)^2},
\end{array}
\end{equation}
typically called the conformal cross ratios or the anharmonic ratios. As a result any analytic function of these ratios $f(u,v)$ satisfies the required symmetry properties and just on symmetry alone we can not find a unique expression for the four-point function. What can be done is to apply the OPE to any two pairs of operators within the four-point function, the result is a double sum over two-point functions. Due to the diagonalization of the two-point function each term in one sum will only connect to terms in the other sum that are from the same conformal family (i.e. descend from the same primary). 
\begin{equation}
\begin{array}{l}
\contraction{}{O_1}{(x_1)}{O_2}
\contraction{O_1(x_1)O_2(x_2)}{O_2}{(x_3)}{O_1}
\langle O_1(x_1)O_2(x_2)O_3(x_3)O_4(x_4)\rangle= \frac{1}{(x_1-x_2)^{\frac{1}{2}(\Delta_1+\Delta_2)}(x_3-x_4)^{\frac{1}{2}(\Delta_3+\Delta_4)}}\left(\frac{x_2-x_4}{x_1-x_4}\right)^{\frac{1}{2}(\Delta_1-\Delta_2)}\\\\\left(\frac{x_1-x_4}{x_1-x_3}\right)^{\frac{1}{2}(\Delta_3-\Delta_4)}\sum_{k} c_{12k}c_{34k}\,G(\frac{1}{2}(\Delta_k-\Delta_1+\Delta_2-s),\frac{1}{2}(\Delta_k-\Delta_3+\Delta_4-s),\Delta_k;u,v),
\end{array}
\end{equation}
where the sum over $k$ represents a sum over all primaries. For all purposes in this paper all operators that appear in the four-point functions are chosen to be scalar operators. Since the raising operator of the conformal algebra acts as a derivative on the operators, it makes sense to group all contributions from a single conformal family into a single function $G(u,v)$, the conformal partial waves or conformal blocks. For a very long time series representations for these conformal blocks have been known\cite{Dobrev1, Dobrev2}, but somewhat more recently under certain conditions closed-from expressions for the conformal blocks have been found \cite{DolanOsborn1},\cite{DolanOsborn2},\cite{DolanOsborn3}.

\begin{figure}[h]
	\centering
		\includegraphics[scale=0.8]{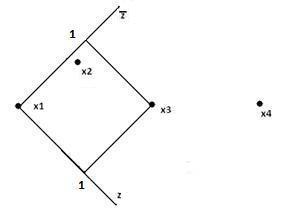}
	\caption{In this picture the coordinate system is displayed in which $z$ and $\bar{z}$ take on a more clear meaning as lightcone coordinates. The point $x_1$ is located at (0,0), $x_3$ at (1,1) and $x_4$ at (L,L). The point $x_2$ is free and is considered at $z\rightarrow0$, $\bar{z}\rightarrow1$.}
	\label{coordinatepicture}
\end{figure}

By use of the coordinate transformations 
\begin{equation}
\begin{array}{ll}
u=z\bar{z}\,,&v=(1-z)(1-\bar{z}),
\end{array}
\end{equation}
they were able to solve the eigenfunctions of the Casimir operator of the conformal algebra. These lightcone coordinates have a simple interpretation if you use conformal transformations to map three out of the four points onto a straight line, this is demonstrated in figure \ref{coordinatepicture}. On an euclidian plane these coordinates play the role of the usual complex coordinates $z=x-iy$ and $\bar{z}=x+iy$. On a Minkowski plane they can be interpreted as lightcone coordinates, in this paper the points are chosen be on a Minkowski plane. These eigenfunctions are of the form 
\begin{equation}
\begin{array}{l}
G_{\Delta_k,s}(z,\bar{z})=\frac{(z\bar{z})^{\frac{1}{2}(\Delta_k-s)}}{z-\bar{z}}[(-\frac{1}{2}z)^sz\,_2F_1(\frac{1}{2}(\Delta_k-\Delta_1+\Delta_2-s-2),\frac{1}{2}(\Delta_k-\Delta_3+\Delta_4-s-2),\Delta_k-s-2,\bar{z})\\\\\times\,_2F_1(\frac{1}{2}(\Delta_k-\Delta_1+\Delta_2+s),\frac{1}{2}(\Delta_k-\Delta_3+\Delta_4+s),\Delta+s,z)-z\leftrightarrow\bar{z}]
\end{array}
\label{conformalblock1}
\end{equation}
These functions because they are the eigenfunctions of the conformal Casimir are by construction conformal invariants and eigenfunctions of the scale invariance generator. As a result they can be interpreted as a function representation of the primaries that appear in the OPE. Also as such the conformal partial wave decomposition given above is analogous to decomposing functions into spherical harmonics.

\subsection{The conformal bootstrap}
\begin{figure}[h]
	\centering
		\includegraphics{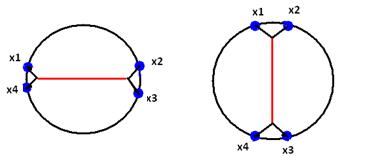}
	\caption{The difference between the t-channel (left) and s-channel (right) is in the way the pairs of operators in the four-point function have been contracted by taking the OPE's}
	\label{channelspic}
\end{figure}

In the previous section a certain choice has been made in the way the operators in the four-point function have been contracted. But the way pairs of operators have been selected in the four-point function should be in some sense arbitrary. If the insertion locations of the operators has been chosen such that in different choices of contraction both OPEs would converge than the result of evaluating the four-point functions should be equal. This is represented in a graphical way in figure \ref{channelspic}, here we distinguish between the s-channel CPW decomposition and the t-channel decomposition. The relation between these two will play a critical role in the upcoming sections.

This conformal bootstrap or crossing symmetry places, as shall be seen, very heavy restrictions on a consistent CFT, for a very long time it has been suggested that it might be strong enough to find a general solution for a CFT \cite{difrancesco}, since as mentioned before all of the dynamic information of a CFT is contained in its operator spectrum and it's OPE coefficients. 

\section{A short review of \cite{KZ}.}
This section will contain a very short review on the argument developed in \cite{KZ} where the authors have derived expressions for the anomalous dimension of operators with conformal dimension $2\Delta+s$ for large spin\footnote{To be fair; that paper gives a more general result than that and demonstrates that if operators with dimension $\Delta_1$ and $\Delta_2$ exist there should exist large spin operators with dimension arbitrarily close to $\Delta_1+\Delta_2$. The special case $\Delta_1=\Delta_2$ will be reviewed here for its lighter notation}, assuming the existence of a conformal primary with dimension $\Delta$. Or to put it more explicitly, their result is that if a unitary CFT contains a scalar operator of dimension $\Delta$ then at large $s$ a tower of double trace operators exists with conformal dimension given by
\begin{equation}
\tau=2\Delta-\frac{2c_{\tau_m}}{s^{\tau_m}},
\label{DoubleTrace}
\end{equation}
where $\tau$ is the conformal twist defined as $\tau=\Delta_k-s$ \cite{Alday2}. The coefficient of the anomalous dimension is given by
\begin{equation}
c_{\tau_m}=f^2\frac{\Gamma(\tau_m+2s_m)}{(-2)^{s_m}\Gamma(\frac{\tau_m+2s_m}{2})^2}\frac{\Gamma(\Delta)^2}{\Gamma(\Delta-\frac{\tau_m}{2})^2},
\label{ADcoef}
\end{equation}
where the subscript $m$ designates that it is a property of the primary operator in the spectrum with lowest twist (excluding the identity operator). The constant $f$ designates the coefficient of the minimal twist operator that appears in the OPE of $O(x)$ with itself. The first observation that has to be made is that the operator product expansion on a Minkowski background due to dimensional analysis has the form of equation (\ref{OPE})
\begin{equation}
O_1(x)O_2(0)=\sum_{O^{(k)}} \frac{c_{12k}}{|x|^{\Delta_1+\Delta_2-\Delta_k}}\frac{x^{\mu_1}...x^{\mu_2}}{|x|^s}O^{(k)}_{\mu_1...\mu_s}(x),
\end{equation}
where the sum is over the operators in the spectrum of the CFT. 

The argument now depends on the evaluation of the four-point function
\begin{equation}
\mathcal{F}(x_1,x_2,x_3,x_4)=\langle O(x_1)O(x_2)O(x_3)O(x_4)\rangle,
\end{equation}
where conformal transformations have been applied to map all the points onto a plane. The points are chosen to be at the lightcone coordinates $x_1=(0,0), x_2=(z,\bar{z}), x_3=(1,1)$ and $x_4=(L,L)$ where $L$ is a large constant in the sense that to any arbitrary precision $|x_1-x_4|\approx|x_2-x_4|\approx|x_3-x_4|$, see figure \ref{coordinatepicture}. As was mentioned in the previous section two different conformal partial wave decompositions can be found; firstly the s-channel where one first takes the OPE between $O(x_2)$ and $O(x_1)$. which results in 
\begin{equation}
\mathcal{F}(z,\bar{z})=\left(\frac{1}{z\bar{z}}\right)^\Delta \sum_{\Delta_k}\sum_{s=0}^\infty c_{s,\Delta_k} G_{s,\Delta_k}(z,\bar{z}).
\end{equation}
where $G_{s,\Delta}(z,\bar{z})$ are the conformal blocks associated to the double-trace operators and their descendants that occur in the OPE and are found by applying equation (\ref{conformalblock1})\footnote{Note that the factor $\frac{1}{2^s}$ in equation (\ref{conformalblock1}) has been suppressed in equations (\ref{ConformalBlock}) and (\ref{CollinearBlock}), if you wish to keep it just remember to multiply the CPW coefficients with a factor $2^s$} to our situation:
\begin{equation}
G_{\Delta,s}=(-1)^s\frac{z\bar{z}}{z-\bar{z}}\left[z^{\frac{\tau+2s}{2}}\bar{z}^{\frac{\tau-2}{2}}\,_2F_1(\frac{\tau+2s}{2},\frac{\tau+2s}{2},\tau+2s,z) \,_2F_1(\frac{\tau-2}{2},\frac{\tau-2}{2},\tau-2,\bar{z})+z\leftrightarrow\bar{z}\right].
\label{ConformalBlock}
\end{equation}
By taking the limit $z\rightarrow0$ the last expression is to leading order given by the collinear blocks
\begin{equation}
g_{\Delta,s}\approx z^{\frac{\tau}{2}}\bar{z}^{\frac{\tau+2s}{2}}\,_2F_1(\frac{\tau+2s}{2},\frac{\tau+2s}{2},\tau+2s,\bar{z})
\label{CollinearBlock}
\end{equation}
Similarly when we take the OPE between $O(x_2)$ and $O(x_3)$ then the resulting decomposition will be designated as the t-channel:
\begin{equation}
\mathcal{F}(z,\bar{z})=\left(\frac{1}{(1-z)(1-\bar{z})}\right)^\Delta \sum_{\Delta_k}\sum_{s=0}^\infty c_{s,\Delta_k} g_{s,\Delta_k}(1-z,1-\bar{z}).
\end{equation}

The distinction between these channels was displayed graphically in figure (\ref{channelspic}). It is important now to note that in the limit $z\rightarrow0$ and $\bar{z}\rightarrow1$ the t-channel conformal blocks reduce to
\begin{equation}
g_{\Delta,s}(1-z,1-\bar{z})\rightarrow -\frac{\Gamma(\tau+2s)}{(-2)^s\Gamma(\frac{\tau+2s}{2})^2}(1-\bar{z})^{\frac{\tau}{2}}(\log(z)+O(1))
\end{equation}
At this point the constraint of unitarity comes in, a unitary CFT places a contraint on the dimensions of the operator spectrum of the CFT, specifically $\tau\geq d-2$ for operators with spin greater then 1 and $\Delta\geq\frac{d-2}{2}$ for scalar operators. There is one exception in the form of the identity operator which has scaling dimension equal to 0. The result is that in a CFT that lives in $d>2$ there exists a so-called twist gap i.e. there exists a certain operator with minimal twist $\tau_m$. Hence by taking the limit $\bar{z}\rightarrow1$ we can keep the conformal blocks related to the identity and the minimal twist operator and neglect the rest of the terms in the CPW decomposition\footnote{In other words in the regime we are interested in we can think of the CFT as one where the only operator that couples to the operators $O$ is the minimal twist operator.} since they are supressed by higher powers of $(1-\bar{z})$. The result is the following approximation for the t-channel CPW decomposition:
\begin{equation}
\mathcal{F}(z,\bar{z})=\left(\frac{z\bar{z}}{(1-z)(1-\bar{z})}\right)^\Delta\left(1-f^2\frac{\Gamma(\tau_m+2s_m)}{(-2)^{s_m}\Gamma(\frac{\tau_m+2s_m}{2})^2} (1-\bar{z})^{\frac{\tau_m}{2}}\log(z)+...\right),
\end{equation}
where $f$ is the OPE-coefficient related to the minimal twist operator in the OPE between $O(x_2)$ and $O(x_3)$. Note that the minimal twist operator does not have to be unique. If the theory under consideration contains multiple primaries with conformal dimension $\tau_m$ then we simply add up their respective conformal blocks.

A key point is that both channels can be made to converge simultaneously by taking $x_2$ to $\bar{z}=1$ and $z=0$ since at this point the lightcones of both $x_1$ and $x_3$ intersect therefore $x_2$ can be made arbitrarily close to both points simultaneously. This leads to the following crossing equation
\begin{equation}
\sum_{\Delta_k}\sum_{s=0}^\infty c_{s,\Delta_k} g_{s,\Delta_k}(z,\bar{z})=\left(\frac{z\bar{z}}{(1-z)(1-\bar{z})}\right)^\Delta\left(1-f^2\frac{\Gamma(\tau_m+2s_m)}{(-2)^{s_m}\Gamma(\frac{\tau_m+2s_m}{2})^2} (1-\bar{z})^{\frac{\tau_m}{2}}\log(z)+...\right).
\label{Crossing}
\end{equation}
At this point it can be argued that the operator spectrum of the CFT needs to contain double-trace operators of conformal dimension close to $2\Delta+s$, since it is the only set of collinear blocks (\ref{CollinearBlock}) that can reproduce the leading order dependance on $z$ of the right-hand side.

\subsection{Evaluating the s-channel CPW decomposition}
To demonstrate that the anomalous dimension can be argued to be of the form given in (\ref{DoubleTrace}) start with the decomposition into collinear blocks of the free field solution of the four point function
\begin{equation}
\left(\frac{1}{z\bar{z}}\right)^\Delta \sum_{s=0}^\infty c_{s,\Delta} g_{s,\Delta}(z,\bar{z})=(z\bar{z})^{-\Delta}+1+\left((1-z)(1-\bar{z})\right)^{-\Delta}.
\end{equation}
In the region we are interested in only one term is projected out
\begin{equation}
\sum_{s=0}^\infty c_{s,\Delta} g_{s,\Delta}(z,\bar{z})=\left(\frac{z\bar{z}}{(1-z)(1-\bar{z})}\right)^\Delta,
\label{LeadingGFT}
\end{equation}
which is exactly the contribution of the identity operator to the crossing equation (\ref{Crossing}). The coefficients of the CPW decomposition are known and have been calculated in for instance \cite{FKcoef} to be:
\begin{equation}
c_{N,s}=\frac{(1+(-1)^s)(\Delta-\frac{d}{2}+1)_N^2(\Delta)_{N+s}^2}{\Gamma(s+1)\Gamma(N+1)(s+\frac{d}{2})_N(2\Delta+N-d+1)_N(2\Delta+2N+s)_s(2\Delta+N+s-\frac{d}{2})_N},
\label{cpwcoef}
\end{equation}
where $d$ is the number of spacetime dimensions and where $(a)_b$ denotes the Pochhammer symbol, which is defined as $(a)_b=\Gamma(a+b)/\Gamma(a)$. At the moment we are interested only in the leading $N=0$ terms, in which case the coefficients reduce to:
\begin{equation}
c_{s}=(1+(-1)^s)\frac{\Gamma(\Delta+s)^2\Gamma(2\Delta+s-1)}{\Gamma(s+1)\Gamma(\Delta)^2\Gamma(2\Delta+2s-1)}.
\end{equation} 
Plugging the collinear blocks from equation (\ref{CollinearBlock}) into the decomposition (\ref{LeadingGFT}) and replacing the sum by an integral over $s$ leads to the expression
\begin{equation}
\int_{0}^\infty ds \, c_{s,\Delta} z^{\Delta}\bar{z}^{\Delta+s}\,_2F_1(\Delta+s,\Delta+s,2\Delta+2s,\bar{z}),
\end{equation}
note that this means that we implicitly expect the sum over large spins to be most relevant. First apply an integral representation of the confluent hypergeometric function to obtain the expression 
\begin{equation}
\int_0^\infty ds \,\int_0^1 dt\, 2\frac{\Gamma(\Delta+s)^2\Gamma(2\Delta+s-1)}{\Gamma(s+1)\Gamma(\Delta)^2\Gamma(2\Delta+2s-1)}\frac{\Gamma(2\Delta+2s)}{\Gamma(s+\Delta)^2}\frac{\left(\frac{\bar{z}t(1-t)}{1-\bar{z}t}\right)^{s+\Delta}}{t(1-t)}.
\end{equation}
Next perform the coordinate transformations:
\begin{equation}
\begin{array}{l}
z=e^{-2\beta},\\
\bar{z}=1-e^{-2\sigma},\\
\hat{s}=s+\Delta.
\end{array}
\end{equation}
This results in the expression
\begin{equation}
\int_0^1 dt\, \int_\Delta^\infty d\hat{s}\, 2\frac{\Gamma(\Delta+\hat{s}-1)\Gamma(2\hat{s})}{\Gamma(\hat{s}-\Delta+1)\Gamma(\Delta)^2\Gamma(2\hat{s}-1)}\frac{\left(\frac{(1-e^{-2\sigma})t(1-t)}{1-(1-e^{-2\sigma})t}\right)^{\hat{s}}}{t(1-t)}.
\end{equation}
The quotient of gamma-functions can be simplified significantly by applying in turn the recurrence relation $\Gamma(x+1)=x\Gamma(x)$, the Stirling approximation $\Gamma(x)=\sqrt{\frac{2\pi}{x}}e^{x\log(x)-x}(1+O(\frac{1}{x}))$ and only keeping the leading order term in an expansion in $\frac{1}{\hat{s}}$. This all results in
\begin{equation}
\frac{4}{\Gamma(\Delta)^2}\int_0^1 dt\, \int_{\Delta}^{\infty} d\hat{s}\, \frac{\hat{s}^{(2\Delta-1)}\left(\frac{(1-e^{-2\sigma})t(1-t)}{1-(1-e^{-2\sigma})t}\right)^{\hat{s}}}{t(1-t)}
\end{equation}
The next point is to determine which spins dominate the above integral. To do so perform a saddle point analysis. Rewriting numerator in the integral as an exponent leads to the dominant spin
\begin{equation}
\hat{s}_0=-\frac{2\Delta-1}{\log\left(\frac{(1-e^{-2\sigma})t(1-t)}{1-(1-e^{-2\sigma})t}\right)}
\end{equation}
Resulting expression is dependant on $t$. To eliminate the $t$-dependance first use the Laplace method to eliminate the integral over $s$ and then find the saddle point for the integral over $t$. The first step leads to
\begin{equation}
\frac{4e^{1-2\Delta}\sqrt{\pi}(2\Delta-1)^{2\Delta-\frac{1}{2}}}{\Gamma(\Delta)^2}\int_0^1 dt\, \frac{\left(-\log\left(\frac{(e^{2\sigma}-1)(1-t)t}{e^{2\sigma}(1-t)+t}\right)\right)^{-2\Delta}}{t(1-t)}
\end{equation}
Doing a series approximation around $(1-t)<<1$ keeping the resulting three singular terms and determining the dominant contribution to $t$ leads to the approximate form
\begin{equation}
t_0=1-\sqrt{\frac{2\Delta-1}{2\Delta+1}}e^{-\sigma}+O(e^{-2\sigma}).
\end{equation}
Now that we got this, the value for $s$ that dominates this sum can be expressed as
\begin{equation}
\log(s_0)=\sigma+\log\left(\frac{(2\Delta-1)^{\frac{3}{2}}(2\Delta+1)^{\frac{1}{2}}}{4\Delta}\right)+O(e^{-\sigma}).
\end{equation}
This demonstrates that an anomalous dimension of the form shown in (\ref{DoubleTrace}) will accurately reproduce the $\bar{z}$ dependance of the subleading term on the right-hand side of the crossing equation (\ref{Crossing}).

\subsection{The anomalous dimension coefficient}
In this section the deriviation of the coefficient (\ref{ADcoef}) will be reviewed. First look at the crossing equation (\ref{Crossing}), the important aspect is that the only place where $\tau$ occurs and the anomalous dimension would not be negligable compared to $s$ is in the power of $z$. Performing the series expansion $z^{\Delta+\frac{\gamma_s}{2}}=z^\Delta(1+\frac{1}{2}\log(z)\gamma_s+O(\gamma_s^2))$ demonstrates that the this will reproduce the factor $\log(z)$ in the subleading term on the right-hand side. This gives a good indication that this approach is on the right track. To recreate the right $\bar{z}$ dependance of the minimal twist block, make the ansatz that $\gamma_s=-2\frac{c_{\tau_m}}{s^{\tau_m}}$. Doing this expansion in the s-channel CPW decomposition and equating the resulting subleading terms in (\ref{Crossing}) results in
\begin{equation}
\begin{array}{l}
\sum_s^\infty -c_s \frac{c_{\tau_m}}{s^{\tau_m}}z^\Delta\log(z)\bar{z}^{\Delta+s}\,_2F_1(\Delta+s,\Delta+s,2\Delta+2s) \\=\left(\frac{z\bar{z}}{(1-z)(1-\bar{z})}\right)^\Delta f^2\frac{\Gamma(\tau_m+2s_m)}{(-2)^{s_m}\Gamma(\frac{\tau_m+2s_m}{2})^2} (1-\bar{z})^{\frac{\tau_m}{2}}\log(z)
\end{array}
\end{equation}  
Taking on the right-hand side only the leading $z^{\Delta}$ part in the region $z<<1$ and ignoring all powers of $\bar{z}$ results in the following defining equation for $c_{\tau_m}$.
\begin{equation}
c_{\tau_m}=f^2\frac{\Gamma(\tau_m+2s_m)}{(-2)^{s_m}\Gamma(\frac{\tau_m+2s_m}{2})^2}\frac{(1-\bar{z})^{\frac{\tau}{2}-\Delta}}{\sum_s^\infty c_s s^{-\tau_m}\,_2F_1(\Delta+s,\Delta+s,2\Delta+2s,\bar{z})}
\label{ctaudef}
\end{equation}
Therefore the only thing left to do is to evaluate the sum in the denominator
\begin{equation}
\sum_s^\infty c_s s^{-\tau_m}\,_2F_1(\Delta+s,\Delta+s,2\Delta+2s,\bar{z}).
\label{DenomSum}
\end{equation}
To do that first perform the transformation $\Delta+s=\frac{A}{\sqrt{\epsilon}}$. Replace the sum over $s$ by an integral over $A$
\begin{equation}
\int_0^\infty \frac{dA}{\sqrt{\epsilon}}\, 2\frac{\Gamma(\frac{A}{\sqrt{\epsilon}})^2\Gamma(\frac{A}{\sqrt{\epsilon}}+\Delta-1)}{\Gamma(\frac{A}{\sqrt{\epsilon}}-\Delta+1)\Gamma(\Delta)^2\Gamma(\frac{2A}{\sqrt{\epsilon}}-1)}(\frac{A}{\sqrt{\epsilon}}-\Delta)^{-\tau_m}\,_2F_1(\frac{A}{\sqrt{\epsilon}},\frac{A}{\sqrt{\epsilon}},\frac{2A}{\sqrt{\epsilon}},1-\epsilon)
\end{equation}
where the final argument of the hypergeometric function is due to the just established approximate relation $s\approx\frac{1}{\sqrt{1-\bar{z}}}$. Next take the limit $\epsilon\rightarrow0$, such as to exploit the previous assertion that it is the large spin contribution that dominates the sum. In this limit the hypergeometric function behaves as 
\begin{equation}
\,_2F_1(\frac{A}{\sqrt{\epsilon}},\frac{A}{\sqrt{\epsilon}},\frac{2A}{\sqrt{\epsilon}},1-\epsilon)\rightarrow \frac{A}{\sqrt{\epsilon}}\frac{4^{\frac{A}{\sqrt{\epsilon}}}}{\epsilon^{\frac{1}{4}}}K_0(2A),
\end{equation}
where $K_0(x)$ is the modified Bessel function of the second kind. Using this and applying the Stirling approximation for the gamma-functions in the coefficient results in the following form for (\ref{DenomSum})
\begin{equation}
\frac{4}{\Gamma(\Delta)^2\epsilon^{\Delta-\frac{\tau}{2}}}\int_0^\infty dA\,A^{2\Delta-\tau_m-1}K_0(2A).
\end{equation}
Applying the following integral identity of the Bessel function $\int_0^\infty dx\,x^{2c-1}K_0(2x)=\frac{1}{4}\Gamma(c)^2$ the previous integral can be solved to give
\begin{equation}
\frac{4}{\Gamma(\Delta)^2\epsilon^{\Delta-\frac{\tau}{2}}}\int_0^\infty dA\,A^{2\Delta-\tau_m-1}K_0(2A)=\frac{1}{\epsilon^{\Delta-\frac{\tau}{2}}}\frac{\Gamma(\Delta-\frac{1}{2}\tau_m)^2}{\Gamma(\Delta)^2}.
\end{equation}
Finally replacing the sum in (\ref{ctaudef}) by this expression results in equation (\ref{ADcoef}) that was quoted at the start of this section:
\begin{equation}
c_{\tau_m}=f^2\frac{\Gamma(\tau_m+2s_m)}{2^{s_m}\Gamma(\frac{\tau_m+2s_m}{2})^2}\frac{\Gamma(\Delta)^2}{\Gamma(\Delta-\frac{\tau_m}{2})^2}.
\end{equation}

\section{Generalizing to arbitrary $N$}
\begin{figure}
	\centering
		\includegraphics[scale=0.8]{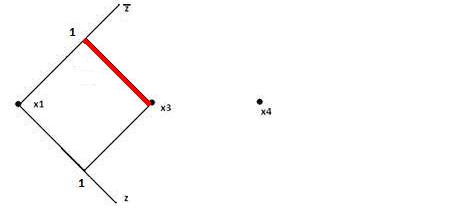}
	\caption{The region where we will apply the conformal bootstrap is marked with the thick red line, compare this with the point in figure \ref{coordinatepicture}}.
	\label{AllOrderCoordinates}
\end{figure}

In this section we will build on from the previously stated results by applying the techniques developed in the paper \cite{KZ} to show that operators have to exist in the spectrum with twist close to $2\Delta+2N$ where $N$ is any positive integer. The result is an expression for the general anomalous dimension of these large spin operators. To do this make the stronger assumption that the crossing equation from before:
\begin{equation} 
\sum_{\Delta_k}\sum_{s=0}^\infty c_{s,\Delta_k} G_{\Delta_k,s}(z,\bar{z})=\left(\frac{z\bar{z}}{(1-z)(1-\bar{z})}\right)^\Delta\left(1-
f^2G_{\tau_m+s_m,s_m}(1-z,1-\bar{z})+...\right),
\end{equation}
not only holds true in the limit $z\rightarrow0$ but also at finite but small separation from 0. To do that we will evaluate the crossing equation to all orders in $z$ and not just the leading part, see figure (\ref{AllOrderCoordinates}). In the end we will find the anomalous dimenion by comparing all the parts of the crossing equation that will contribute logarithmically in $z$.

\subsection{The full t-channel}
First we will see what the full $z$-dependance of the t-channel will contribute. To do that we will do a full series expansion of the identity operator contribution and collect all the parts of the minimal twist block that are proportional to $\log(z)$. To do this first we will first look at the t-channel conformal block and only keep the leading part in the limit $\bar{z}\rightarrow1$
\begin{equation}
g_{\Delta,s}(1-z,1-\bar{z})=(1-z)^{\frac{\Delta+s}{2}}(1-\bar{z})^{\frac{\Delta-s}{2}}\,_2F_1(\Delta+s,\Delta+s,\frac{\Delta+s}{2},1-z),
\end{equation}
The result for the t-channel will be
\begin{equation}
\begin{array}{l}
\mathcal{T}=\frac{1}{(1-\bar{z})^\Delta}\left(\sum_{q=0}^\infty \frac{\Gamma(\Delta+q)}{\Gamma(\Delta)\Gamma(q+1)}z^{\Delta+q}\right)\\\\-f^2\left(\sum_{w=0}^\infty \frac{\Gamma(\Delta-\frac{\tau_m+2s_m}{2}+w)}{\Gamma(\Delta-\frac{\tau_m+2s_m}{2})\Gamma(w+1)}z^{\Delta+w}\right)(1-\bar{z})^{\frac{\tau_m}{2}-\Delta}\log(z)\sum_{u=0}^\infty\left(\frac{\Gamma(\frac{\tau_m+2s_m}{2}+u)}{\Gamma(\frac{\tau_m+2s_m}{2})\Gamma(u+1)}z^u\right)^2,
\end{array}
\end{equation}
where all terms of the minimal twist block expansion that do not contain a factor $\log(z)$ have been omitted. This double sum in the minimal twist block can be made more practical by reorganizing the terms and sorting them by powers $z$:
\begin{equation}
\begin{array}{l}
\mathcal{T}=\frac{1}{(1-\bar{z})^\Delta}\left(\sum_{q=0}^\infty \frac{\Gamma(\Delta+q)}{\Gamma(\Delta)\Gamma(q+1)}z^{\Delta+q}\right)\\\\
-f^2(1-\bar{z})^{\frac{\tau_m}{2}-\Delta}\log(z) \sum_{q=0}^\infty z^{\Delta+q} \sum_{u=0}^q \frac{\Gamma(\Delta-\frac{\tau_m+2s_m}{2}+q-u)}{\Gamma(\Delta-\frac{\tau_m+2s_m}{2})\Gamma(q-u+1)} \left(\frac{\Gamma(\frac{\tau_m+2s_m}{2}+u)}{\Gamma(\frac{\tau_m+2s_m}{2})\Gamma(u+1)}\right)^2,
\end{array}
\end{equation}

\subsection{The full s-channel}
Unfortunately the collinear blocks (\ref{CollinearBlock}) that have been used so far are only the leading order term that one obtains when expanding the full conformal blocks (\ref{ConformalBlock}) in powers of $z$ therefore if we want to take the right-hand side at finite separation we have to be fair and also take the full $z$ dependance of the s-channel conformal blocks into account. Therefore to find the anomalous dimension for every value of $N$ the left-hand side needs to be fully expanded in powers of $z$ and both sides of the equation have to matched term by term.

First focus on the full left hand side of the above equation
\begin{equation}
\begin{array}{l}
\sum_{\Delta_k}\sum_{s=0}^\infty c_{s,\Delta_k} G_{\Delta_k,s}(z,\bar{z})= \\\\\sum_{\Delta_k}\sum_{s=0}^\infty c_{s,\Delta_k}( (-1)^s\frac{z\bar{z}}{z-\bar{z}}
[z^{\Delta+N+s}\bar{z}^{\Delta+N-1}\,_2F_1(\Delta+N+s,\Delta+N+s,2(\Delta+N+s),z) \\\\\times\,_2F_1(\Delta+N-1,\Delta+N-1,2(\Delta+N-1),\bar{z})+z\leftrightarrow\bar{z}]),
\end{array}
\label{confblockdecomp}
\end{equation}
where the conformal dimension of the double trace operators has already been applied. The defining sum representation of the hypergeometric function is given by
\begin{equation}
\,_2F_1(a,b,c,x)=\sum_{n=0}^\infty \frac{(a)_n (b)_n}{(c)_n}\frac{z^n}{n!}
\end{equation}
Therefore by expanding the prefactor $\frac{z\bar{z}}{z-\bar{z}}$ in (\ref{confblockdecomp}) in series of $z$ and multiplying this with a series expansion of the hypergeometric series that have $z$ as an argument a full power series of in $z$ can be found:
\begin{equation}
\begin{array}{l}
\sum_N \sum_s \sum_k -c_{s,n}(-1)^s \frac{z^{k+1}}{\bar{z}^k}(\bar{z}^{\Delta+N-1}\,_2F_1(\Delta+N-1,\Delta+N-1,2\Delta+2N-2,\bar{z})\\\\ \sum_l \frac{\Gamma(\Delta+N+s+l)^2\Gamma(2\Delta+2N+2s)}{\Gamma(\Delta+N+s)^2\Gamma(l+1)\Gamma(2\Delta+2n+2s+l)}z^{\Delta+N+s+l}- \bar{z}^{\Delta+N+s}\,_2F_1(\Delta+N+s,\Delta+N+s,2\Delta+2N+2s,\bar{z})\\\\\sum_{l'=0}^\infty \frac{\Gamma(\Delta+N+l'-1)^2\Gamma(2\Delta+2N-2)}{\Gamma(\Delta+N-1)^2\Gamma(l'+1)\Gamma(2\Delta+2N+l'-2)}z^{\Delta+N-1+l'})\\
\end{array}
\end{equation}
These two quadruple sums can be reordered into a single quadruple sum in the following way

\begin{equation}
\begin{array}{l}
\sum_N \sum_s \sum_k c_{s,n}(-1)^s\left(-\frac{z^{k+1}}{\bar{z}^k}\right)\sum_{l=0}^\infty (\bar{z}^{\Delta+N-1}\theta_{s+1}(l)\,_2F_1(\Delta+N-1,\Delta+N-1,2\Delta+2N-2,\bar{z})\\\\
\frac{\Gamma(\Delta+N+l-1)^2\Gamma(2\Delta+2N+2s)}{\Gamma(\Delta+N+s)^2\Gamma(l-s)\Gamma(2\Delta+2N+s+l-1)} - \bar{z}^{\Delta+N+s}\,_2F_1(\Delta+N+s,\Delta+N+s,2\Delta+2N+2s,\bar{z})\\\\ \frac{\Gamma(\Delta+N+l-1)^2\Gamma(2\Delta+2N-2)}{\Gamma(\Delta+N-1)^2\Gamma(l+1)\Gamma(2\Delta+2N+l-2)})z^{\Delta+N-1+l}\\
\end{array}
\end{equation}
Where the step function $\theta_k(l)$ is defined as:
\begin{equation}
\theta_k(l)=\left\{
\begin{array}{c}
0\,\textrm{if}\, l<k\\
1\,\textrm{if}\, l\geq k\\
\end{array}\right.
\end{equation}
The mayor simplifying argument is that if we keep the power of $z$ fixed then $l$ is also fixed, but as was shown in \cite{KZ} it is known that the sum over $s$ in the CPW decomposition will be dominated by the high spin terms. The left term in the sum over $l$ is exactly zero though if $s+1>l$ therefore if doing a saddle point analysis over the sum over $s$ is justified then it should be safe to ignore the entire left most term. This would be nice since the only difference between the rightmost term for every index pair $(l,k)$ and the collinear block is a factor that although large is independant of $s$. Therefore if it was not the case already that the terms ignored are negligable then we could just take $\bar{z}$ closer to 1 such that the saddle-point of the important points gets shifted to larger values. 

This is a very heuristic argument, but doing the actual sum over all spins over the terms that are not neglected (as is done in appendix B) demonstrates that this sum will scale as $\frac{1}{1-\bar{z}}^\Delta$, and therefore reproduces the scaling of the identity operator contribution to the right-hand side. The finite number of terms that have been neglected have their scaling with variation in $\bar{z}$ determined by the hypergeometric functions $\,_2F_1(\Delta+N-1,\Delta+N-1,2\Delta+2N-2,\bar{z})$. Which  will to leading order diverge logarithmically in the limit $\bar{z}\rightarrow1$. Therefore these terms only form logarithmic corrections and can safely be ignored.   

Also doing a parameter transformation $q=N+K+l$ and eliminating $l$ in favor of $q$ leads to the expression
\begin{equation}
\begin{array}{l}
\sum_{q=0}^\infty \sum_{N=0}^q \sum_{k=0}^{q-N}\sum_{s=0}^\infty (-1)^s c_{s,N} \frac{\Gamma(\Delta+q-k-1)^2\Gamma(2\Delta+2N-2)}{\Gamma(\Delta+N-1)^2\Gamma(q-N-k+1)\Gamma(2\Delta+N+q-k-2)}\\\\\times\,_2F_1(\Delta+N+s,\Delta+N+s,2\Delta+2N+2s,\bar{z})z^{\Delta+q}
\end{array}
\label{fullschannel}
\end{equation}
The nice intermediate result is that you have an explicit sum over the powers of $z$ so now you can equate the conformal bootstrap equation term-by-term. Also notice that the problem: how many terms have the same power of $z$. Is equivalent to the problem in how many ways can I distribute $q$ identical objects over 3 distinct containers. Therefore the number of terms with the same power is equal to 
$\begin{pmatrix}2+q\\q \end{pmatrix}=\frac{1}{2}(q+1)(q+2)$.

\subsection{The anomalous dimension}
As before introduce an anomalous dimension in the exponent of $z$ of the form $\gamma(N,s)=-\frac{2c_{\tau_m}(N)}{s^{\tau_m}}$ note that it is anticipated that the anomalous dimension coefficient will depend on $N$. This anomalous dimension will recreate the logarithmic dependance on the right hand side via the expansion $z^{\Delta+\frac{\gamma(N,s)}{2}}=z^\Delta(1+\frac{1}{2}\log(z)\gamma(N,s)+O(\gamma^2)$). This results in crossing equation of the form
\begin{equation}
\begin{array}{l}
\sum_{q=0}^\infty \sum_{N=0}^q \sum_{k=0}^{q-N}\sum_{s=0}^\infty (-1)^s c_{s,N} \frac{\Gamma(\Delta+q-k-1)^2\Gamma(2\Delta+2N-2)}{\Gamma(\Delta+N-1)^2\Gamma(q-N-k+1)\Gamma(2\Delta+N+q-k-2)}\\\\\times\,_2F_1(\Delta+N+s,\Delta+N+s,2\Delta+2N+2s,\bar{z})z^{\Delta+q}\left(1-\log(z)\frac{c_{\tau_m}(N)}{s^{\tau_m}}\right)\\\\=

\frac{1}{(1-\bar{z})^\Delta}\left(\sum_{q=0}^\infty \frac{\Gamma(\Delta+q)}{\Gamma(\Delta)\Gamma(q+1)}z^{\Delta+q}\right)\\\\
-f^2(1-\bar{z})^{\frac{\tau_m}{2}-\Delta}\log(z) \sum_{q=0}^\infty z^{\Delta+q} \sum_{u=0}^q \frac{\Gamma(\Delta-\frac{\tau_m+2s_m}{2}+q-u)}{\Gamma(\Delta-\frac{\tau_m+2s_m}{2})\Gamma(q-u+1)} \left(\frac{\Gamma(\frac{\tau_m+2s_m}{2}+u)}{\Gamma(\frac{\tau_m+2s_m}{2})\Gamma(u+1)}\right)^2.
\end{array}
\end{equation}
To reiterate this is why we only need the terms of the minimal twist block that contain factors of $\log(z)$. The anomalous dimension in the exponent of $z$ in the s-channel is the only place where factors like $\log(z)$ can manifest. Our lives are made a lot simpler if we only equate these parts, rather than match the entire extension of the crossing equation.
\begin{equation}
\begin{array}{l}
\sum_{N=0}^{q}\sum_{k=0}^{q-N}\sum_{s=0}^\infty (-1)^{s+1} c_{s,N} \frac{\Gamma(\Delta+q-k-1)^2\Gamma(2\Delta+2N-2)}{\Gamma(\Delta+N-1)^2\Gamma(q-N-k+1)\Gamma(2\Delta+N+q-k-2)}\\\\\times\,_2F_1(\Delta+N+s,\Delta+N+s,2\Delta+2N+2s,\bar{z})z^{\Delta+q}\log(z)\frac{c_{\tau_m(N)}}{s^{\tau_m}}\\\\
=\,-\left(\frac{\bar{z}}{(1-\bar{z})}\right)^\Delta  z^{\Delta+q}f^2\frac{\Gamma(\tau_m+2s_m)}{(-2)^{s_m}\Gamma(\frac{\tau_m+2s_m}{2})^2} (1-\bar{z})^{\frac{\tau_m}{2}}\log(z)\sum_{u=0}^q \frac{\Gamma(\Delta-\frac{\tau_m+2s_m}{2}+q-u)}{\Gamma(\Delta-\frac{\tau_m+2s_m}{2})\Gamma(q-u+1)} \left(\frac{\Gamma(\frac{\tau_m+2s_m}{2}+u)}{\Gamma(\frac{\tau_m+2s_m}{2})\Gamma(u+1)}\right)^2.
\end{array}
\label{Ncrossing}
\end{equation}
The sum over $s$ in equation (\ref{Ncrossing}) can be resolved with the assumption that it is the terms at large $s$ that dominate the sum. The calculation is done in appendix A. The result is:
\begin{equation}
\begin{array}{l}
\sum_{N=0}^q \sum_{k=0}^{q-N}c_{\tau_m}(N)\frac{\Gamma(\Delta+q-k-1)^2\Gamma(2\Delta+2N-2)}{\Gamma(\Delta+N-1)^2\Gamma(q-N-k+1)\Gamma(2\Delta+N+q-k-2)}\\\\\times\alpha(N)(1-\bar{z})^{\frac{1}{2}\tau_m-\Delta}\Gamma(\Delta-\frac{1}{2}\tau_m)^2z^{\Delta+q},
\end{array}
\end{equation}
where the factor $\alpha(N)$ is defined as
\begin{equation}
\alpha(N)\equiv\frac{\Gamma(\Delta+N-1)^2\Gamma(2\Delta+N-3)}{(\Delta-1)^2\Gamma(\Delta-1)^4\Gamma(2\Delta+2N-3)}.
\end{equation}
The result is the following crossing equation for the first subleading term
\begin{equation}
\begin{array}{l}
\sum_{N=0}^q \sum_{k=0}^{q-N}c_{\tau_m}(N)\frac{\Gamma(\Delta+q-k-1)^2\Gamma(2\Delta+2N-2)}{\Gamma(\Delta+N-1)^2\Gamma(q-N-k+1)\Gamma(2\Delta+N+q-k-2)}\\\\\times\alpha(N)\Gamma(\Delta-\frac{1}{2}\tau_m)\, =\,  f^2\frac{\Gamma(\tau_m+2s_m)}{(-2)^{s_m}\Gamma(\frac{\tau_m+2s_m}{2})^2} \sum_{u=0}^q \frac{\Gamma(\Delta-\frac{\tau_m+2s_m}{2}+q-u)}{\Gamma(\Delta-\frac{\tau_m+2s_m}{2})\Gamma(q-u+1)} \left(\frac{\Gamma(\frac{\tau_m+2s_m}{2}+u)}{\Gamma(\frac{\tau_m+2s_m}{2})\Gamma(u+1)}\right)^2.
\end{array}
\end{equation}

From this point onwards all the anomalous dimension coefficients $c_{\tau_m}(N)$ can be found recursively. To do this take the term related to the greatest value of $N$ (i.e. $c_{\tau_m}(q)$) and pull it out of the sum. The result is
\begin{equation}
\begin{array}{l}
c_{\tau_m}(q)=\frac{1}{\alpha(q)\Gamma(\Delta-\frac{1}{2}\tau)^2}[\sum_{u=0}^q \frac{\Gamma(\Delta-\frac{\tau_m+2s_m}{2}+q-u)}{\Gamma(\Delta-\frac{\tau_m+2s_m}{2})\Gamma(q-u+1)} \left(\frac{\Gamma(\frac{\tau_m+2s_m}{2}+u)}{\Gamma(\frac{\tau_m+2s_m}{2})\Gamma(u+1)}\right)^2 \\\\\times

f^2\frac{\Gamma(\tau_m+2s_m)}{(-2)^{s_m}\Gamma(\frac{\tau_m+2s_m}{2})^2}-\sum_{N=0}^{q-1}\sum_{k=0}^{q-N}c_{\tau_m}(N)\frac{\Gamma(\Delta+q-k-1)^2\Gamma(2\Delta+2N-2)\alpha(N)\Gamma(\Delta-\frac{1}{2}\tau_m)^2}{2\Gamma(\Delta+N-1)^2\Gamma(q-N-k+1)\Gamma(2\Delta+q+N-k-2)}].
\end{array}
\end{equation}
Reinstating the cumbersome factors $\alpha(N)$ leads to 
\begin{equation}
\begin{array}{l}
c_{\tau_m}(q)=\frac{\Gamma(2\Delta+2q-3)\Gamma(\Delta-1)^2\Gamma(\Delta)^2\Gamma(q+1)}{\Gamma(\Delta+q-1)^2\Gamma(2\Delta+q-3)\Gamma(\Delta-\frac{1}{2}\tau)^2}[\sum_{u=0}^q \frac{\Gamma(\Delta-\frac{\tau_m+2s_m}{2}+q-u)}{\Gamma(\Delta-\frac{\tau_m+2s_m}{2})\Gamma(q-u+1)} \left(\frac{\Gamma(\frac{\tau_m+2s_m}{2}+u)}{\Gamma(\frac{\tau_m+2s_m}{2})\Gamma(u+1)}\right)^2 
\\\\\times f^2\frac{\Gamma(\tau_m+2s_m)}{(-2)^{s_m}\Gamma(\frac{\tau_m+2s_m}{2})^2}-\sum_{N=0}^{q-1}\sum_{k=0}^{q-N}c_{\tau_m}(N)\frac{\Gamma(\Delta+q-k-1)^2(2\Delta+2N-3)\Gamma(2\Delta+N-3)\Gamma(\Delta-\frac{1}{2}\tau_m)^2}{\Gamma(2\Delta+N+q-k-2)\Gamma(q-N-k+1)\Gamma(\Delta-1)^2\Gamma(\Delta)^2\Gamma(N+1)}].
\end{array}
\label{anomdimdef}
\end{equation}Which trivially reduces to the result of \cite{KZ},\cite{FKPSD} in the case $q=0$. To keep the interpretation clear, the one positive term in square brackets is the contribution from the anomalous dimension of the conformal block $G_{2\Delta+2q+s,s}$ the negative terms are corrections due to resumming over the subleading terms of the conformal blocks $G_{2\Delta+2N+s,s}$ where $N<q$.

\section{An AdS point of view} 
In \cite{penedones1},\cite{penedones2},\cite{penedones3},\cite{maldacena2} the four-point amplitude of four points located at the boundary of AdS was considered. It was found in \cite{tHooft} that in the large energy small scattering angle kinematical regime the scattering amplitude is dominated by disconnected graphs and crossed ladder diagrams in which gravitons are exchanged. By doing an s-channel conformal partial wave decomposition of the disconnected graphs and assuming that the rest of the amplitude can be created by designing an anomalous dimension that reproduces the tree-level diagrams. 

By this method they conclude that it is the limiting behaviour of the anomalous dimension in the regime $s>>N>>1$ should be given by 
\begin{equation}
\gamma(N,s)=-\frac{4G}{\pi L^3}\frac{N^4}{s^2}.
\end{equation}
Assuming that it is graviton exchange that dominates the four-point function. Here $G$ is the gravitational constant and $L$ plays the role of the radius of AdS.\footnote{Note that in equation (4.4) of \cite{penedones2} the authors have chosen to omit a factor $\frac{1}{2}$, this is why in table (\ref{GravityCorrespondence}) $c_{\tau_m}$ and not $2c_{\tau_m}$ is displayed.}. 

We will now demonstrate that the results from the previous section seem too agree with their results. We can do this by making the dual assumption that it is the stress-energy tensor that plays the role of the minimal twist conformal block. In this case the OPE coefficient $f$ of $T_{\mu\nu}$ with $O$ twice is given in \cite{KZ} by
\begin{equation}
f^2=\frac{4\Delta^2}{45\pi} \frac{G}{L^3}
\label{GravityPrediction}
\end{equation}
\begin{table}[h]
\begin{tabular}{|l|l|l|l|}
\hline
N&$\tilde{c}_{\tau_m}(N)$&&\\
\hline
&$\Delta=3$&$\Delta=4$&$\Delta=\frac{9}{2}$\\
\hline
0&6&24&$\frac{1323}{32}$\\
\hline
1&42&120&$\frac{5859}{32}$\\
\hline
2&156&720&$\frac{16299}{32}$\\
\hline
3&420&840&$\frac{36099}{32}$\\
\hline
4&930&1680&$\frac{69483}{32}$\\
\hline
5&1806&3024&$\frac{121443}{32}$\\
\hline
6&3192&5040&$\frac{197739}{32}$\\
\hline
7&5256&7920&$\frac{304899}{32}$\\
\hline
8&8190&11880&$\frac{450219}{32}$\\
\hline
9&12210&17160&$\frac{641763}{32}$\\
\hline
10&17556&24024&$\frac{888363}{32}$\\
\hline
11&24492&32760&$\frac{1199619}{32}$\\
\hline
12&33306&43680&$\frac{1585899}{32}$\\
\hline
13&44310&57120&$\frac{2058339}{32}$\\
\hline
14&57840&73440&$\frac{2628843}{32}$\\
\hline
15&74256&93024&$\frac{331083}{32}$\\
\hline
16&93942&116280&$\frac{4115499}{32}$\\
\hline
17&117306&143640&$\frac{5059299}{32}$\\
\hline
18&144780&175560&$\frac{6156459}{32}$\\
\hline
19&176820&212520&$\frac{7422723}{32}$\\
\hline
20&213906&255024&$\frac{8874603}{32}$\\
\hline
\end{tabular}
\caption{The first 21 reduced anomalous dimension coefficients given by $\tilde{c}_{\tau_m}(N)=\frac{\pi L^3 c_{\tau_m}(N)}{4G}$ for some values of $\Delta$.}
\label{GravityCorrespondence}
\end{table}

\begin{figure}[h!]
\centering
	\includegraphics[width=.7\linewidth]{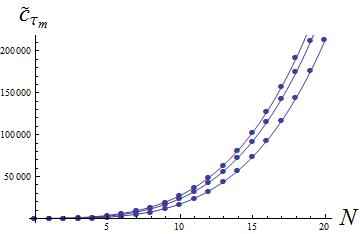}
  \label{fig:sub1}
\caption{A plot of the reduced anomalous dimension coefficients in table (\ref{GravityCorrespondence}). The solid lines are the exact 4th-order polynomials that match the data, see eqs (\ref{polyfit}), (\ref{polyfit2}) and (\ref{polyfit3}).}
\label{fig:test}
\end{figure}
Note that therefore both the AdS energy shifts and the CFT anomalous dimension are both proportional to $\frac{4G}{\pi L^3}$ so we will suppress this factor when comparing the two. In table (\ref{GravityCorrespondence}) the anomalous dimension coefficients are given for some values of $\Delta$. The N-dependance of these numbers is clarified in the below plot. One can always construct a unique $(n-1)$th-order polynomial that fits exactly through $n$ points. By multiplying with an appropriate inverse Vandermonde matrix, we find the unique 20th-order polynomials that fit exactly through the sets of 21 data points in table (\ref{GravityCorrespondence}). For the choice $\Delta=3$ this polynomial is found to be
\begin{equation}
\tilde{c}_{\tau_m}(N)=N^4+6N^3+14N^2+15N+6,
\label{polyfit}
\end{equation}
or equivalently
\begin{equation}
c_{\tau_m}(N)=\frac{4G}{\pi L^3}\left(N^4+6N^3+14N^2+15N+6\right),
\end{equation}
where all coefficients of terms with powers of $N$ greater than 4 are exactly zero. The coefficient of the leading term in the limit $N>>1$ of this polynomial is identical to the predicted asymptotic behaviour of (\ref{GravityPrediction}) in AdS, demonstrating that there is a very satisfactory correspondence between these results. In the original submission only this leading order term was extracted and compared to result in \cite{maldacena2}, after publication an expression for the anomalous dimension coefficients was derived in \cite{Kaviraj} for $\tau_{m}=s_{m}=2$ the closer fitting procedure presented in this version that leads to (\ref{polyfit}) shows complete agreement with the result obtained therein.

Similarly for the other values for $\Delta$ found in table (\ref{GravityCorrespondence}) the exact polynomials that fit through the anomalous dimensions for $\Delta=4$ is given by
\begin{equation}
c_{\tau_m}(N)=\frac{4G}{\pi L^3}\left(N^4+10N^3+35N^2+50N+24\right),
\label{polyfit2}
\end{equation}
and for $\Delta=\frac{9}{2}$
\begin{equation}
c_{\tau_m}(N)=\frac{4G}{\pi L^3}\left(N^4+12N^3+\frac{197}{4}N^2+\frac{159}{2}N+\frac{1323}{32}\right).
\label{polyfit3}
\end{equation}
Which can also be shown to correspond with later results found in \cite{Kaviraj}\footnote{After posting it was pointed out by the authors of \cite{Alday1} that the anomalous dimension coefficient (\ref{anomdimdef}) in this paper match results obtained earlier for $\mathcal{N}=4$ SYM with operators in the four-point function that have conformal dimension $\Delta=4$. To check this the relevant CFT data is that the stress-energy tensor plays the role of the minimal twist operator hence $\tau_{m}=s_{m}=2$ and the relevant OPE coefficient is given by $f^{2}=\frac{2\Delta^{2}}{45n^{2}}=\frac{32}{45n^{2}}$ where $n$ is proportional to the amount of degrees of freedom.}.
\section{Conclusion}
It has been shown that the argument for additivity in twist space for operators with very large spin previously obtained in \cite{KZ},\cite{FKPSD} generalizes to all the double-trace operators that are required to construct the t-channel identity operator that appears in the OPE of two scalar operators $O(x)$. These have the symbolic form $O\partial^s\Box^NO$. An exact expression for the anomalous dimension of these operators has been derived and appears to be supported by dual gravitational calculations.

The required assumptions i.e. a twist gap and crossing symmetry are very general and therefore applicable to a large class of CFTs including strongly coupled theories, unlike past results this method does not assume a large $N$ limit. It neither makes use of gravitational duals, unlike what is done in a.o. \cite{hmr},\cite{penedones1},\cite{penedones2}. Rather any unitary CFT has to satisfy these conditions to maintain consistency with crossing symmetry. The anomalous dimension of these operators obtained earlier in \cite{KZ},\cite{FKPSD} for the specific $N=0$ case can be extended to apply to double trace operators with arbitrary $N$.

This has only been demonstrated explicitly for $d=4$. It seems safe to conjecture that the points in twist space $2\Delta+2N$ limit points for all unitary CFTs that live in $d>2$, but as there exist no closed-form expressions for the conformal blocks in arbitrary dimension this method does not trivially generalize to arbitrary dimensions.

The implications of this result are quite powerful on the AdS side. It is an open question what kind of demands a CFT has to satisfy in order for it to have a dual that can be interpreted as a gravitational particle theory, this is discussed in for instance \cite{FKnorm},\cite{Kyriakos1},\cite{Heemskerk}. It turns out that all CFTs that live in 4D and satisfy unitarity have a sector of operators with very large spin that resembles a holographic CFT. The reason for this is that when applying the AdS dictionary to the operator $O$ we get a field $\phi$, but additionally we obtain a set of operators in the spectrum of the CFT of which the gravitational duals can be interpreted as the elements of the Fock space of the field $\phi$. 

An interesting question in this line of thinking is when and how does this mechanism break down. On the AdS side the anomalous dimension of a double trace operator can be interpreted as the gravitational interaction potential between two bound objects. From this picture it is clear that lowering the spin at some point will bring the objects close enough together that the interaction potential becomes non-negligable.

\section*{Acknowledgments}
I am grateful for the support from my thesis advisor Kyriakos Papadodimas, the large amount of discussions with whom on CFTs and AdS/CFT have proven to be very useful over the course of this research.

\appendix



\section*{Appendix A: Solving the sum over all spins}
\addcontentsline{toc}{section}{Appendix A: Solving the sum over all spins}
In \cite{KZ} a method was developed to approximate the sum over all spins that occurs in the s-channel comformal partial wave decomposition in what is essentially the $N=0$ case. This appendix will demonstrate that this method with minor alterations can be applied to resolve the sum for arbitrary values of $N$ as well. Remember the part of equation (\ref{fullschannel}) that looked like
\begin{equation}
\sum_{s=0}^\infty c_{s,N}\,_2F_1(\Delta+N+s,\Delta+N+s,2\Delta+2N+2s,\bar{z}),
\end{equation}
where the CPW coefficients are for $d=4$ given in \cite{FKcoef} by
\begin{equation}
c_{N,s}=\frac{(1+(-1)^s)(\Delta-1)_N^2(\Delta)_{N+s}^2}{\Gamma(s+1)\Gamma(N+1)(s+2)_N(2\Delta+N-3)_N(2\Delta+2N+s)_s(2\Delta+N+s-2)_N}.
\end{equation}
First pull out all the factors that are independant of spin
\begin{equation}
\alpha(N)\equiv\frac{\Gamma(\Delta+N-1)^2\Gamma(2\Delta+N-3)}{\Gamma(N+1)\Gamma(\Delta-1)^2\Gamma(\Delta)^2\Gamma(2\Delta+2N-3)},
\end{equation}
and apply the integral form of the hypergeometric function that was applied in section 3
also replace the sum over spins by an integral, the resulting expression is
\begin{equation}
\begin{array}{l}
\alpha(N)\int_{0}^\infty ds \, \frac{(s+1)(2\Delta+2N+s-1)\Gamma(2\Delta+N+s-2)(2\Delta+2N+2s-1)}{\Gamma(s+N+2)}\\\\\times\int_0^1 dt\, \frac{\left(\frac{(1-e^{-2\sigma})t(1-t)}{1-(1-e^{-2\sigma})t}\right)^{\Delta+N+s}}{t(1-t)}
\end{array}
\label{appendixintegral1}
\end{equation}
We know that it is the large spin terms that will dominate this sum in the case where $N=0$. The location of the saddle point turns out to be independant of $N$, to show this. Do the parameter transformation $\hat{s}=\Delta+N+s$ and apply the Stirling approximation $\Gamma(x)=\sqrt{\frac{2\pi}{x}}\left(\frac{x}{e}\right)^x\left(1+O(\frac{1}{x})\right)$ to the gamma functions. 
The result is that to leading order in $\frac{1}{s}$
\begin{equation}
\frac{(s+1)(2\Delta+2N+s-1)\Gamma(2\Delta+N+s-2)(2\Delta+2N+2s-1)}{\Gamma(s+N+2)}\propto s^{2\Delta-1},
\end{equation}
which is similar to equation (B.6) from \cite{KZ}, therefore from this point onwards the analysis performed there carries over step-by-step. Resulting in the same conclusion that the dominant contribution to the sum over all spins is given by terms corresponding to
\begin{equation}
s_0=\frac{1}{\sqrt{1-\bar{z}}}.
\end{equation}

\subsubsection*{Resolving the sum}
For this reason it is sensible to make the ansatz that the $\bar{z}$ on the right hand side of equation (\ref{Ncrossing}) is recreated on the left-hand side by an anomalous dimension of the form $-\frac{2c_{\tau_m}(N)}{s^{\tau_m}}$. Since this factor will only significantly connect to the dominant part in the  sum over all spin, and therefore recreate the $(1-\bar{z})^{\frac{\tau_m}{2}}$ dependance of the minimal twist block. To apply this knowledge perform another parameter substitution $\hat{s}=\frac{A}{\sqrt{\epsilon}}$ to equation (\ref{Ncrossing}), where $\epsilon$ is related to $\bar{z}$ via $\epsilon=1-\bar{z}$.
\begin{equation}
\begin{array}{l}
\int_0^\infty \frac{dA}{\sqrt{\epsilon}}\,c_{\tau_m}(N)\left(\frac{A}{\sqrt{\epsilon}}\right)^{-\tau_m}\,\frac{\alpha(N)\Gamma(\frac{A}{\sqrt{\epsilon}})^2\Gamma(\frac{A}{\sqrt{\epsilon}}-N-\Delta+2)\Gamma(\frac{A}{\sqrt{\epsilon}}+\Delta+N)\Gamma(\frac{A}{\sqrt{\epsilon}}+\Delta-2)}{\Gamma(\frac{A}{\sqrt{\epsilon}}-\Delta-N-1)\Gamma(\frac{A}{\sqrt{\epsilon}}-\Delta+2)\Gamma(\frac{2A}{\sqrt{\epsilon}})\Gamma(\frac{A}{\sqrt{\epsilon}}+\Delta+N-2)}\\\\\times\,_2F_1(\frac{A}{\sqrt{\epsilon}},\frac{A}{\sqrt{\epsilon}},\frac{2A}{\sqrt{\epsilon}},1-\epsilon)
\end{array}
\end{equation}
Now apply the Stirling approximation and once again only keep the leading term in $\epsilon\rightarrow0$
\begin{equation}
\alpha(N)c_{\tau_m}(N)\int_0^\infty \frac{dA}{\sqrt{\epsilon}}2^{-2\frac{A}{\sqrt{\epsilon}}+2}\sqrt{\pi}\left(\frac{A}{\sqrt{\epsilon}}\right)^{2\Delta-\frac{3}{2}-\tau_m}\,_2F_1(\frac{A}{\sqrt{\epsilon}},\frac{A}{\sqrt{\epsilon}},\frac{2A}{\sqrt{\epsilon}},1-\epsilon).
\end{equation}
Taking the limit $\epsilon\rightarrow0$ we can replace the hypergeometric function with equation (B.17) from \cite{KZ}
\begin{equation}
\,_2F_1(\frac{A}{\sqrt{\epsilon}},\frac{A}{\sqrt{\epsilon}},\frac{2A}{\sqrt{\epsilon}},1-\epsilon)\rightarrow \sqrt{\frac{A}{\pi}}\frac{4^{\frac{A}{\sqrt{\epsilon}}}}{\epsilon^{\frac{1}{4}}}K_0(2A),
\end{equation}
where $K_0(x)$ is the modified Bessel function of the second kind. Plugging this approximation into the expression above leads to
\begin{equation}
\begin{array}{l}
\alpha(N)c_{\tau_m}(N)\int_{0}^\infty \frac{dA}{\sqrt{\epsilon}} 2^{-2\frac{A}{\sqrt{\epsilon}}+2}\sqrt{\pi}\left(\frac{A}{\sqrt{\epsilon}}\right)^{2\Delta-\frac{3}{2}-\tau_m}\sqrt{\frac{A}{\sqrt{\pi}}}\frac{4^{\frac{A}{\sqrt{\epsilon}}}}{\epsilon^{\frac{1}{4}}}K_0(2A)\\\\ = 4\alpha(N)c_{\tau_m}(N)\epsilon^{\frac{1}{2}\tau_m-\Delta}\int_0^\infty dA A^{2(\Delta-\frac{1}{2}\tau_m)-1}K_0(2A)\\\\=\alpha(N)c_{\tau_m}(N)\epsilon^{\frac{1}{2}\tau_m-\Delta}\Gamma(\Delta-\frac{1}{2}\tau_m)^2
\end{array}
\label{Sintegral}
\end{equation}
Where final step was an application of the integral identity $\int_0^\infty dx \, x^{2y-1}K_0(2x)=\frac{1}{4}\Gamma(y)^2$. This result can be directly applied to solve the sum over $s$ in equation (\ref{Ncrossing}).

\section*{Appendix B: Recreating the identity operator contribution}
\addcontentsline{toc}{section}{Appendix B: Recreating the identity operator contribution}
Sometimes it is good to check and see whether what you are doing makes sense. To this end it is important to see whether or not equation (\ref{fullschannel}) in section 4:
\begin{equation}
\begin{array}{l}
\mathcal{F}(z,\bar{z})=\sum_{q=0}^\infty \sum_{N=0}^q \sum_{k=0}^{q-N}\sum_{s=0}^\infty (-1)^s c_{s,N} \frac{\Gamma(\Delta+q-k-1)^2\Gamma(2\Delta+2N-2)}{\Gamma(\Delta+N-1)^2\Gamma(q-N-k+1)\Gamma(2\Delta+N+q-k-2)}\\\\\times\,_2F_1(\Delta+N+s,\Delta+N+s,2\Delta+2N+2s,\bar{z})z^{\Delta+q},
\end{array}
\end{equation}
still reproduces the identity operator contribution of the t-channel. First resolve the sum over all spins, as was done in appendix A only this time without the anomalous dimension (i.e. take $c_{\tau_m}(N)=0$):
\begin{equation}
\mathcal{F}(z,\bar{z})=\sum_{q=0}^\infty\sum_{N=0}^q\sum_{k=0}^{q-N} \frac{\Gamma(\Delta+q-k-1)^2(2\Delta+2N-3)\Gamma(2\Delta+N-3)}{\Gamma(q-N-k+1)\Gamma(2\Delta+q+N-k-2)\Gamma(\Delta-1)^2\Gamma(N+1)}\frac{1}{(1-\bar{z})^\Delta}z^{\Delta+q}.
\end{equation}
From this point on we have the proper scaling with $\bar{z}$. This is consistent with the identity operator contribution to the t-channel. To have full consistency the following statement needs to be true
\begin{equation}
\sum_{N=0}^q\sum_{k=0}^{q-N} \frac{\Gamma(\Delta+q-k-1)^2(2\Delta+2N-3)\Gamma(2\Delta+N-3)}{\Gamma(q-N-k+1)\Gamma(2\Delta+q+N-k-2)\Gamma(\Delta-1)^2\Gamma(N+1)}=\frac{\Gamma(\Delta+q)}{\Gamma(\Delta)\Gamma(q+1)}.
\label{mysterysum}
\end{equation} 
A proof for this statement is left open, though interpreting the right hand side as a binomial coefficient and applying the Chu-Vandermonde identity twice seems to be fruitful. This constraint does check out analytically for all values of $q$ up to 10.



\begin{thebibliography}{99}

\bibitem{KZ} 
  Z.~Komargodski and A.~Zhiboedov,
  JHEP {\bf 1311}, 140 (2013)
  [arXiv:1212.4103 [hep-th]].

\bibitem{FKPSD} 
  A.~L.~Fitzpatrick, J.~Kaplan, D.~Poland and D.~Simmons-Duffin,
  JHEP {\bf 1312}, 004 (2013)
  [arXiv:1212.3616 [hep-th]].

\bibitem{Alday2} 
  L.~F.~Alday and J.~M.~Maldacena,
  JHEP {\bf 0711}, 019 (2007)
  [arXiv:0708.0672 [hep-th]].

\bibitem{Maldacena} 
  O.~Aharony, S.~S.~Gubser, J.~M.~Maldacena, H.~Ooguri and Y.~Oz,
  Phys.\ Rept.\  {\bf 323}, 183 (2000)
  [hep-th/9905111].

\bibitem{RattazziPaper} 
  R.~Rattazzi, V.~S.~Rychkov, E.~Tonni and A.~Vichi,
  JHEP {\bf 0812}, 031 (2008)
  [arXiv:0807.0004 [hep-th]].

\bibitem{Dobrev1}
V.K. Dobrev, V.B. Petkova, S.G. Petrova and I.T. Todorov,
\textit{Dynamical derivation of vacuum operator product expansion in
Euclidean conformal quantum field theory}, Phys. Rev. {\bf D13}
(1976) 887-912.

\bibitem{Dobrev2}
V.K. Dobrev, G. Mack, V.B. Petkova, S.G. Petrova and I.T.
Todorov, {\it Harmonic Analysis on the $n$ - Dimensional
Lorentz Group and Its Applications to Conformal Quantum Field
Theory}, Lecture Notes in Physics, No 63, 280 pages (Springer
Verlag, Berlin-Heidelberg-New York, 1977)
  
\bibitem{penedones1} 
  L.~Cornalba, M.~S.~Costa and J.~Penedones,
  JHEP {\bf 0709}, 037 (2007)
  [arXiv:0707.0120 [hep-th]].

\bibitem{penedones2} 
  L.~Cornalba, M.~S.~Costa, J.~Penedones and R.~Schiappa,
  Nucl.\ Phys.\ B {\bf 767}, 327 (2007)
  [hep-th/0611123].

\bibitem{penedones3} 
  L.~Cornalba, M.~S.~Costa, J.~Penedones and R.~Schiappa,
  JHEP {\bf 0708}, 019 (2007)
  [hep-th/0611122].

\bibitem{maldacena2} 
  X.~O.~Camanho, J.~D.~Edelstein, J.~Maldacena and A.~Zhiboedov,
  arXiv:1407.5597 [hep-th].

\bibitem{tHooft}
	G.~'t Hooft, \textit{Graviton Dominance in Ultrahigh Energy Scattering}, Phys. Lett. \textbf{B198} (1987)
61.

\bibitem{Fitzpatrick:2014vua} 
  A.~L.~Fitzpatrick, J.~Kaplan and M.~T.~Walters,
  JHEP {\bf 1408}, 145 (2014)
  [arXiv:1403.6829 [hep-th]].

\bibitem{FKcoef} 
  A.~L.~Fitzpatrick and J.~Kaplan,
  JHEP {\bf 1210}, 032 (2012)
  [arXiv:1112.4845 [hep-th]].

\bibitem{FKnorm} 
  A.~L.~Fitzpatrick, E.~Katz, D.~Poland and D.~Simmons-Duffin,
  JHEP {\bf 1107}, 023 (2011)
  [arXiv:1007.2412 [hep-th]].


\bibitem{Kyriakos1} 
  S.~El-Showk and K.~Papadodimas,
  JHEP {\bf 1210}, 106 (2012)
  [arXiv:1101.4163 [hep-th]].

\bibitem{Heemskerk} 
  I.~Heemskerk, J.~Penedones, J.~Polchinski and J.~Sully,
  JHEP {\bf 0910}, 079 (2009)
  [arXiv:0907.0151 [hep-th]].

\bibitem{difrancesco}
	P. di Francesco, P. Mathieu and D. Senechal;
	\emph{Conformal Field Theory}, Springer Science \& Business Media, 1997

\bibitem{DolanOsborn1} 
  F.~A.~Dolan and H.~Osborn,
  Nucl.\ Phys.\ B {\bf 678}, 491 (2004)
  [hep-th/0309180].
  
\bibitem{DolanOsborn2} 
  F.~A.~Dolan and H.~Osborn,
  Nucl.\ Phys.\ B {\bf 678}, 491 (2004)
  [hep-th/0309180].

\bibitem{DolanOsborn3} 
  F.~A.~Dolan and H.~Osborn,
  arXiv:1108.6194 [hep-th].

\bibitem{hmr} 
  L.~Hoffmann, L.~Mesref and W.~Ruhl,
  Nucl.\ Phys.\ B {\bf 608}, 177 (2001)
  [hep-th/0012153].

\bibitem{Alday1} 
  L.~F.~Alday, A.~Bissi and T.~Lukowski,
  arXiv:1410.4717 [hep-th].

\bibitem{Kaviraj} 
  A.~Kaviraj, K.~Sen and A.~Sinha,
  arXiv:1502.01437 [hep-th].



\end{thebibliography}
\end{document}